# Controlling the Information Flow in Spreadsheets

Vipin Samar, Sangeeta Patni
Extensio Software, Inc.,
19763 Merritt Dr. Cupertino, CA, USA
{vsamar, spatni}@extensio.com, 1.408.961.6050

**ABSTRACT**

*There is no denying that spreadsheets have become critical for all operational processes including financial reporting, budgeting, forecasting, and analysis. Microsoft® Excel has essentially become a scratch pad and a data browser that can quickly be put to use for information gathering and decision-making. However, there is little control in how data comes into Excel, and how it gets updated. The information supply chain feeding into Excel remains ad hoc and without any centralized IT control. This paper discusses some of the pitfalls of the data collection and maintenance process in Excel. It then suggests service-oriented architecture (SOA) based information gathering and control techniques to ameliorate the pitfalls of this scratch pad while improving the integrity of data, boosting the productivity of the business users, and building controls to satisfy the requirements of Section 404 of the Sarbanes-Oxley Act.*

## 1   INTRODUCTION

There are plenty of examples of full-fledged Microsoft® Excel based business applications with multiple inputs, outputs, and graphs that are being used for critical business processes in many companies. The reasons for Excel's widespread adoption by its 150 million business users include its perceived simplicity, familiarity, and modeling abilities. However, the businesses also have to live with the various data collection and logic errors that creep into these spreadsheets. For the purpose of this paper, we would be focusing mainly on the errors in data collection and in the flow of information in and out of Excel.

While primitive data gathering tools that import external data into Microsoft Excel have been available from the inception, most users still depend upon populating their handpicked data into Excel using *Control-C* and *Control-V*. Given the nature of critical operations that the business users perform with Excel, and the increased control measures mandated by the Sarbanes-Oxley Act, it is increasingly becoming urgent for businesses to remove dependency on *Control-C* and *Control-V* primitives and also address the quality, freshness, and the accuracy of data in Excel.

This paper starts with providing some historical perspective of how users have traditionally managed the data-entry process for Excel. We then discuss some of the common pitfalls of getting data into the Excel spreadsheet, followed by how section 404 of Sarbanes-Oxley exacerbates the problems.

To address the problems associated with the information supply chain of Excel, we propose a Service-Oriented Architecture (SOA) based model coupled with an Excel Add-In. We describe how users can automate data collection through the use of Information Services instead of copying data, and discuss how this boosts productivity, improves data integrity, and addresses some of the requirements of Sarbanes-Oxley Act without compromising the flexibility and ease-of-use of Excel.

## 2  MANAGING INFORMATION INSIDE EXCEL SPREADSHEET

Spreadsheets started off as a personal productivity tool for calculations and managing operations with persistent data.  It then evolved into a scratch pad to keep lists, perform mathematical functions, and show graphs.  Later, Excel began to be used for collecting data from multiple places, building information models, and conducting iterative and incremental analysis.  Somewhere along this time, Excel also started getting used as a group productivity tool even though there were no specific features that helped with the collaborative effort.  Today, Excel based applications exist in most companies performing tasks ranging from simple ones such as weekly reporting to complex ones such as financial accounting, budgeting, forecasting, and operational planning.  Despite huge investments in IT, most business users still depend upon handpicking data for Excel from CRM, ERP, Portals, and databases.

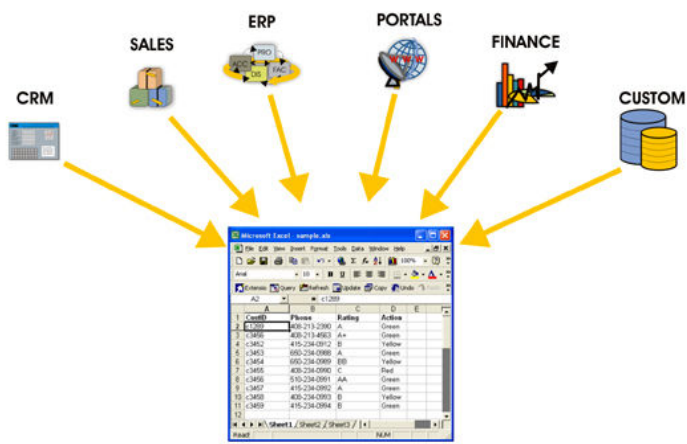

In a survey by CFO IT [Durfee 2004] of 168 finance executives on the use of IT by corporate finance departments, they found that only 2 out of a list of 14 finance-specific technologies were widely used: spreadsheets (100%) and basic budgeting and planning systems (66%).  The survey also reported that when executives were asked about the usage of spreadsheets five years from now, 91% of them thought that the spreadsheets would have the same or more importance.

The extent of Excel usage in enterprises is indeed quite deep and widespread.  According to Forrester's Keith Gile [Gile 2005], *"14 percent of end users are producers - those who create analytic reports and author enterprise reports.  The remaining 86 percent are consumers of the information and data."*  He further adds that most business users (25% of the total) and casual users (30% of the total) prefer canned reports, or reports in Excel formats that they can then parameterize and use.  Further, the extended enterprise users (38%) need read-only Excel reports.

As Excel is here to stay for a long time within businesses, it is important to mitigate some of the pitfalls without sacrificing its ease-of-use.  Excel is indeed a free-format scratch pad, but unfortunately is getting used by business users who are not trained in structured programming, and version control [Panko 2000].  The free world of spreadsheets makes them vulnerable to the following types of errors:
- **Input errors**: These errors are due to inaccurate cut-paste, inadvertent changes in cells, incorrect links, import of incorrect data or import with wrong parameters.
- **Logic errors**: These errors are due to incorrect formulas, and incorrect input data.
- **Usage errors**: These errors include incorrect use of functions, ranges, and references.

For the remainder of this document, we would be concerned mainly about input errors, how data comes into Excel, and how it is maintained in the spreadsheet.

## 2.1 Current Methods for Getting Data into Excel

Here are the drawbacks of the common ways/tools used by users to get data in Excel:

| Tools Used | Limitations |
|---|---|
| Copy-paste external data with Control-C, Control-V | <ul><li>Manual process, with high error and no validation</li><li>No trace of the source of data</li><li>Users need to reformat the data as per their requirements</li><li>Neither systematic nor reproducible process</li></ul> |
| Import the *.xls* or *.csv* file created by the application | <ul><li>No linkage with the originating application</li><li>Users need to reformat the data as per their requirements</li><li>Does not support update of data using Excel</li><li>Limited use when data comes from multiple applications</li></ul> |
| Import from databases using ODBC | <ul><li>Considered complex for typical business users as it requires knowledge of database structures and SQL</li><li>Not applicable when the source is not a database</li><li>Practical only with few resources and few users</li><li>No meta-data available for advanced manipulations</li></ul> |
| Import using web query | <ul><li>Typically used for HTML reports from public pages</li><li>Requires users to specify the complete URL along with parameters and embedded authentication information.</li><li>No ability to select fields, or header information.</li></ul> |
| Import data using web services (*Excel 2003 Prof. edition only*) | <ul><li>Considered complex for business users as they need to specify the web service, and map the XML schema</li><li>Works only with web service resources</li><li>Authentication issues with multiple web services</li></ul> |
| Access data warehouse with Excel Add-in | <ul><li>Works only against historical data in data warehouse</li><li>Not suitable for daily operational needs of business users</li><li>Not possible to update data</li></ul> |

None of the above approaches are satisfactory as they have the following deficiencies:
- **Loss of Linkage.** Once data is copied without tracking the source of data, any updates in the source cannot be propagated to the eco-system.
- **No Support for Multiple Sources.** Much of the data needed comes from multiple types of sources. Doing it manually is error-prone and time consuming.
- **Stale Data**. Without easy ways to refresh data, decisions are made with old data if it takes a long time to collect data.
- **Non-repeatable and difficult-to-manage process**: Each user performs these operations independently on their own spreadsheets making it impossible for IT to verify, automate, audit or validate the data transfer mechanism. Further, due to lack of IT management and control, efforts spent in building expensive data imports/exports cannot be re-used within the organization.
- **Read-Only Data**. Without being able to update the back-end right from Excel, users have to learn new applications, and depend upon copy-paste to transfer data.
- **Lack of Ease-of-Use**. If the data collection process is cumbersome, the users are likely to go back to their current cut-paste model to meet their objectives.
- **Missing Audit Trail**. Without the means to track the spreadsheets as they change, compliance and auditing becomes difficult.

## 2.2 Sarbanes-Oxley (SOX) Enters the Picture

SOX mandates that companies need some controls over all aspects of financial reporting. Most data that business users touch have some impact, whether direct or indirect, into the financial reporting of the business. Starting from a sales person, to an inventory manager, or a warehouse manager, all submit their figures using Excel. The channel or the resellers also aggregate their bookings, returns, and forecasts, all using Excel.

Computer World [Horowitz 2004] reported, *"Fannie Mae made a **$1.2 billion accounting error** last year because of what it called "honest mistakes made in a spreadsheet. TransAlta Corp. took a **$24 million charge** after a bidding snafu caused by a cut-and-paste error in an Excel spreadsheet."*

The real problem is that the *Control-C, Control-V* takes almost zero time to master, and while this works in the favor of busy business users, it doesn't create a systematic and reproducible process that can stand the scrutiny of their departmental policies and financial auditors. This increases the long-term cost of the Excel-based solution including its maintenance, debugging, auditing and compliance.

Excel's autonomy is what made it popular, but without adequate controls, this can become a huge liability. The challenge is to address the following SOX compliance issues while keeping usability into consideration:
- Control over the data input and output process for spreadsheets
- Control over who is authorized to get that data
- Control over who modified the data, and when
- Control over how repeatable is this entire process

## 2.3 Paying for the Inefficient Information Supply-Chain for Excel

The key constituencies that pay for these inherent Excel weaknesses are:
1. **Business Users**: The users pay dearly by the hours they take in data hunting-and-gathering. They also take hours in learning how to get data from multiple sources.
2. **IT Department**: The IT departments pay because once the data leaves their home repositories they lose control over it. As the data spreads through the enterprise, there is loss of the attached metadata, security, management, integrity, and data constraints. Locking the data repositories from user access is not a viable alternative, as it would force IT to do all the reporting work, and increase their load.
3. **Business**: The business suffers because decisions are made with stale, inaccurate, and insufficient data. There is no single version of truth across the enterprise.

## 3 PUTTING CONTROLS ON THE INFORMATION FLOW INTO EXCEL

This section proposes Information Delivery with a Service-Oriented-Architecture (SOA) based server and an Excel add-in for controlling the flow of information in/out of Excel.

There are two root causes for the information flow problem. The first is that the users deal with data directly and touch it by hand, and the second is that there is no control over the data once it leaves the IT managed application repositories and reaches Excel. Also, we believe that if there is any complexity, it should be on the IT side and not the user side, because IT is more equipped and trained (and paid) for this complexity. If the data can be hidden behind an " information service" that the user can pick and use, the user would not have to know the various details on how to get the data. Further, since services would be built, designed and managed by IT, they will provide better data and access controls.

## 3.1 Introduction to Information Services

Information Services, broadly speaking, can be defined as a logically grouped set of information elements, extracted from data source(s) about an information entity (the "key"). An information service may have information elements combined from multiple data source, or could be the result of some transform operation on the original data. For example, a Customer Information service may take a customer ID and return the Customer name, address, phone number, and credit rating. End users view and consume this service without knowing that the customer name, address and phone numbers have come from their back-end CRM application, while the credit rating has come from their custom database application or a SOAP service. These information services are typically hosted on a server within the data-center.

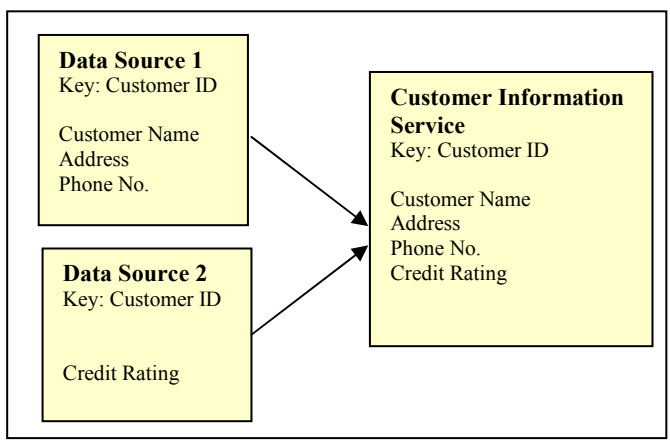

By moving to a service-based approach, the users just specify the service they want (e.g. customer rating service), and they get the related data from the service directly into their Excel cells, without having to know the where, the how, and the when of the data. It is the service now that knows the details of getting the data and not the user.

## 3.2 Introduction to Service-oriented Architecture (SOA)

An information server using service-oriented architecture enables a loosely coupled integration of the back-end resources such that it masks the application consuming this data from the underlying IT complexity. This complexity includes the back-end resource structure, data-format, session management, security, connection pooling, and caching.

The main value of the SOA-based architecture is in deriving value out of the existing applications and resources without duplicating data, business logic or security efforts. The SOA based framework is typically composed of the following elements:
- Native interfaces to connect, authenticate, and access back-end sources. A SOA based system can access heterogeneous resources using SOAP or resource native interfaces such as SQL, HTTP, SAP BAPI, or custom protocol.
- Logic to combine and unify the data from multiple sources. This can be a simple JOIN statement to complex scripting depending upon the user requirements.
- Framework to expose the combined data elements as an information service that can then be consumed by end-users, devices or other applications.
- Framework to publish service directory and service schema

## 3.3 Introduction to Excel Add-in

Microsoft Excel provides well-defined means to add custom commands and features to Excel using Excel Add-ins. Examples include statistical and financial packages. While most Add-ins typically manipulate the Excel data, they can do lot more including contacting an external server and fetching external data. They can also define their own toolbar that can then be used in conjunction with the services offered by the Add-in.

## 3.4 SOA and Excel Add-In for Improving Information Flow Into Excel

The combination of a SOA based Information Delivery Server and an end-user interface coupling technology (Excel Add-in) at the user's desktop provides the foundation for managing the information flow in and out of Excel. The underlying SOA architecture improves the accessibility and quality of data from back-end resources, while the Excel Add-in automates data collection, delivers data to Excel, and limits the human error.

Here are the main requirements of a SOA-based Information flow system for Excel:
- **Automate data collection**. Allow users to collect/refresh data without *copy-paste*
- **Keep data connected**. Enable one version of truth by linking cells to the source.
- **Update data.** Allow users to update data directly from Excel.
- **Audit cell updates.** Allow viewing of previous 'n' values for auditing or tracing.
- **Version Control.** Enables users to easily switch between versions of worksheets.
- **Centralized control.** Centrally manage all service/resource changes and security.
- **Primed for growth**. Support a large user base, both internal and external.

For a SOA based framework to provide data for end-user consumption, the technology needs the following additional components:
- Modules to map the result presentation layer to interfaces such as Excel
- Protocols to connect and authenticate various end-user interfaces
- Interface specific service publication and registry
- Interface specific update process

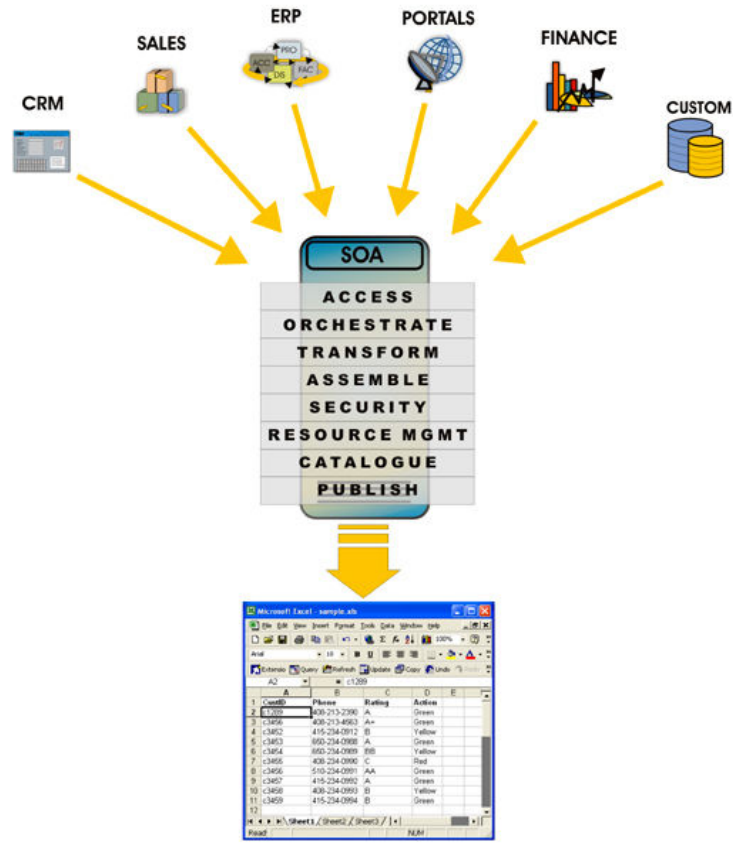

The SOA based information delivery server contains metadata about the services such as the data keys, connection details, transformation, assembly, resource information, and security. The service specific meta-data is stored in a XML rules database and used at run-time for request processing. If there are any changes to the query, they can be made without any user impact.

The server publishes a directory containing the services and attributes. The user chooses the service and input parameters, and sends the request to the server. The server responds by calling the enterprise application using the meta-data, getting the result in the native format, extracting the relevant data, converting the data format and finally sending the data to the end-user.

The Excel Add-in provides the end-user interface and the delivery component. It sends the parameterized service request to the server using XML/HTTP. It then interprets the XML results sent back by the server and updates the cells. By sending the results back as XML, the Information Delivery Server retains the ability to send any other meta-data attached to the cell data. This allows for future extensibility. Examples include suggested refresh period, the associated update services for this element, and any special formatting.

### 3.5 Defining and Building Information Services

The IT department in conjunction with the business analysts and users typically develops the information services. Alternately, the application vendors can ship prepackaged information services that access their own applications, and then the IT department can create information services that combine data from those services. Once the services are defined, the business analysts can select the layout and the fields required by the users.

IT Developers take the following steps to build the services:
1. Identify information entities such as customer, including validation and mapping.
2. Identify the data sources that contain information elements for the entity.
3. Define the rules to fetch/update data from data store using the development tools. The rules can be written using resource specific wizards such as the SQL query wizard, SOAP wizard and the XML/HTML filter.
4. If required, define the rules for transformation, mapping, and assembly of data.
5. Define the presentation layer as appropriate for Excel, desktop, PDA or mobile.
6. Publish the information service on the services directory for the end-user.
7. Assign user/group permissions for information services.

Depending upon the complexity of the service, it may take from a few minutes to a few hours or more to build a set of information services, whether for lookup, reports, update, or alerts. SOA based architecture provides the required infrastructure for IT to serve, manage and control end-user access to information services.

### 3.6 SOA and Excel Add-In Integration Architecture

The overall architecture is shown below.

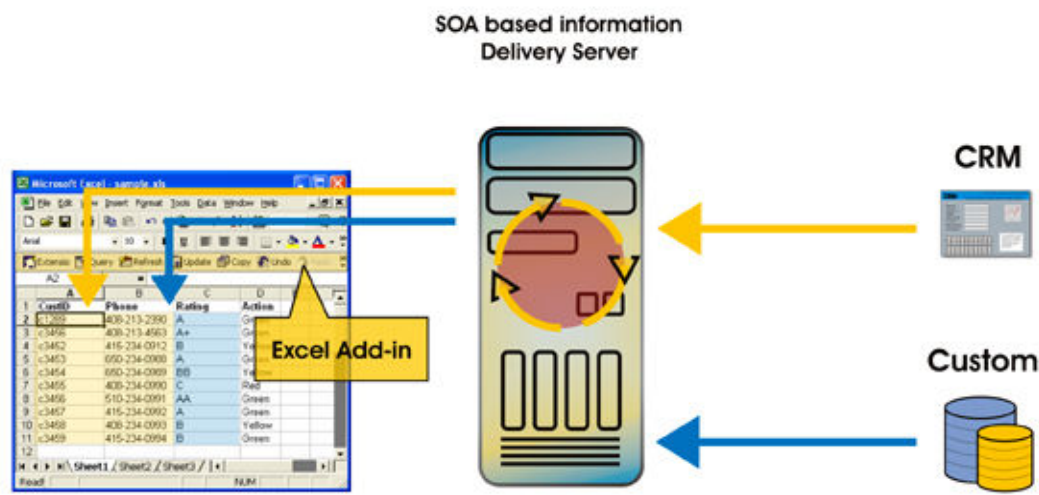

Copyright Extensio Software, Inc., 2005

The Information Delivery Server and Excel Add-in work in tandem to manage the information flow into Excel spreadsheets:

| User Requirements | SOA Contribution | Excel Add-in Contribution |
|---|---|---|
| Easily select data from multiple data stores, and refresh them later. | Provides information services containing elements spanning multiple data-sources. | Provides the ability to link and refresh Excel cells using information services. |
| Eliminate copy-paste, and use of SQL/XML to get data in Excel | Information services that hide the technical details of accessing data-sources | Provides menu based interface to get data directly within Excel |
| Update/Insert data in back-end from Excel | Provides the update interface if supported by applications | Consistent mechanism to specify and send updates |
| Have a replicable data collection process | Centrally available services allow reuse across users | Have persistent links to services in Excel |
| Maintain "single version of truth" and cell-integrity | Live on-demand connection to data-sources ensures latest data | Provides refresh on-demand, audit, version control and cell protection, for cell integrity |
| Ability to add internal and external data sources | Allows addition of information elements from new data sources with the tools | The newly published services are accessible from the service directory |

### 3.7 SOA and Excel Add-In for Data Integrity, Audit, and SOX Compliance

SOA and the Excel Add-In help maintain the integrity and the freshness of the data. Since the data comes directly from the applications in a controlled manner, the integrity of the spreadsheet is not put in risk from copy-pasted data. Further, a direct refresh capability ensures that the financial reports have the latest data. SOA ensures that if there is any error in one of the input elements, it can be easily fixed and then refreshed by the dependent spreadsheets. SOA based Information Delivery Server can also provide centrally managed authorization control on per service-level for users/groups.

The Excel Add-In shows only the authorized services to users. It can also block the users from making any direct changes in the server-supplied data, while permitting for changes made with "refresh". The Add-In can provide audit trail of the cells including the earlier values, time-of-update, and the user making the change. The user also can checkpoint the Excel spreadsheet and restore it to any of the previous checkpoints. This avoids the user having to keep manage multiple versions of the spreadsheets.

Thus, the Information Delivery Server in combination with an Excel Add-In provide controls for SOX compliance, and make the data collection and update process repeatable, verifiable and controllable.

### 4 BENEFITS OF SOA ENABLED INFORMATION DELIVERY FOR EXCEL

For any IT solution to succeed, it is important that it address the requirements of all major constituencies including the end-users, the business and the IT. The end-users on one end want agility, flexibility, and immediate delivery of data, and the IT on the other hand wants control, ease-of-deployment, and reuse of existing investments.

## 4.1 Benefits for the Business User

By moving away from a data-centric model to a service-specific model, the users get the following benefits without having to do any heavy lifting, or using custom tools:
- **Boosts productivity** by eliminating the time spent in moving data in/out of Excel.
- **Makes accurate decisions with latest data** as users can access fresh data.
- **Updates the back-end applications from Excel directly** instead of copy-paste.
- **Provides one version of truth** and reduce proliferation of spreadsheets.
- **Provides audit trail of changes** for tracking and debugging.
- **Reduces dependency on IT staff** for periodic reports as users have direct access.

## 4.2 Benefits for the IT Department

By moving to a SOA based information delivery, the IT benefits significantly:
- **Reduces risks of bad data** by delivering centrally managed Information Services
- **Shrinks the load of creating custom reports** by making users self-sufficient.
- **Connects internal and external applications** to Excel using the SOA model.
- **Support internal and external users** with support for HTTP/HTTPS
- **Reduces compliance issues** by putting easy-to-use controls for integrity.
- **Secures the Information Supply Chain** with centralized control and encryption.
- **Support Excel, desktop applications, and mobile** without any extra effort.

# 5 SUMMARY

Most businesses depend upon spreadsheets for financial reporting and managing operational processes. However, this flexibility comes at a significant cost of data integrity, loss of productivity, and loss of IT control. The paper highlighted some of the reasons for these information flow problems, and proposed a service-oriented architecture through which one can have adequate controls to mitigate these risks.

With a combination of service-oriented architecture (SOA) and Excel Add-In, the users can use information services instead of cutting-and-pasting data. This retains the link of the Excel with the enterprise data sources, and improves IT control over the data in the spreadsheet. Besides helping in compliance with Section 404 under Sarbanes-Oxley, the proposed approach boosts productivity, improves data integrity, and increases security while retaining Excel's flexibility and ease-of-use.